\documentstyle[twoside,fleqn,espcrc2,epsfig]{article}

\title{%
\raisebox{2.5cm}[0pt][0pt]{\makebox[0pt][l]{\parbox{16cm}{\normalsize%
\mbox{}\hfill KANAZAWA 99--17 \\ 
\mbox{}\hfill HUB-EP-99/37 \\ 
\mbox{}\hfill ZIB preprint SC 99--28}}}%
Decorrelation of the topological charge in tempered
Hybrid Monte Carlo simulations of QCD\thanks{Talk
presented by H. St\"uben at Lattice '99, Pisa, Italy.}}

\author{%
E.-M. Ilgenfritz\address{%
Institute for Theoretical Physics,
Kanazawa University, Kanazawa 920--1192, Japan},
W. Kerler\address{%
Institut f\"ur Physik, Humboldt Universit\"at zu Berlin, 10115 Berlin,
Germany}
and H. St\"uben\address{%
Konrad-Zuse-Zentrum f\"ur Informationstechnik Berlin, 14195 Berlin, Germany}
}

\begin{document}

\begin{abstract}
We study the improvement of simulations of QCD with
dynamical Wilson fermions by combining the Hybrid Monte Carlo algorithm 
with parallel tempering.  As an indicator for
decorrelation we use the topological charge.
\end{abstract}

\maketitle
\thispagestyle{empty}

\section{INTRODUCTION}

Decorrelation of the topological charge in Hybrid Monte Carlo (HMC)
simulations of QCD with dynamical fermions is a long standing problem.
For staggered fermions an insufficient tunneling rate of the
topological charge $Q_t$ has been observed \cite{MMP,Pisa}.  For Wilson
fermions the tunneling rate is adequate in many cases \cite{SESAM,RB1}.  
However on large lattices and for large values of $\kappa$ near the 
chiral limit the distribution of $Q_t$ is not symmetric even after more 
than 3000 trajectories (see Figure~1 of \cite{SESAM} and similar observations
by CP-PACS \cite{RB2}).

The idea of parallel tempering is to improve transitions in parameter
regions where tunneling is suppressed by opening ways through parameter
regions with little suppression. In QCD the method has
been applied successfully for staggered fermions \cite{Boyd}. In
\cite{UKQCD} parallel tempering has been used to simulate QCD with
O($a$)-improved Wilson fermions without finding any gain, however, 
with only two ensembles which does not take advantage of the main idea 
of the method.

Here we use parallel tempering in conjunction with HMC to simulate QCD 
with (standard) Wilson fermions in a parameter range close to the critical 
region, where the method is expected to be most advantageous. The gain 
achieved is demonstrated by studying time series and histograms of the
topological charge.

\newcommand{\qtext}[1]{\makebox[7.5cm][c]{\textbf{#1}}}
\newcommand{\qpic}[1]{\epsfig{file=q.#1.eps,width=7.5cm,%
bbllx=0,bblly=0,bburx=568,bbury=220}}

\begin{figure*}[t]
\qtext{Standard HMC}\hspace{1cm}\qtext{Tempered HMC}

\vspace{2ex}

\qpic{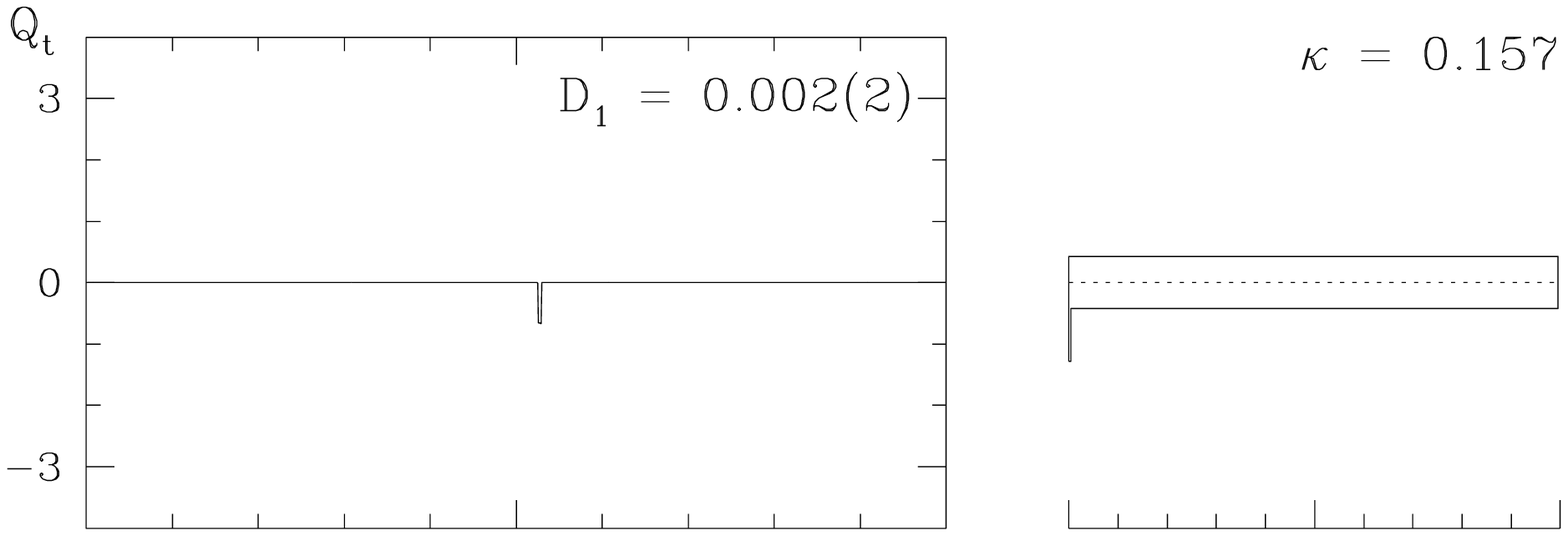}\hspace{1cm}\qpic{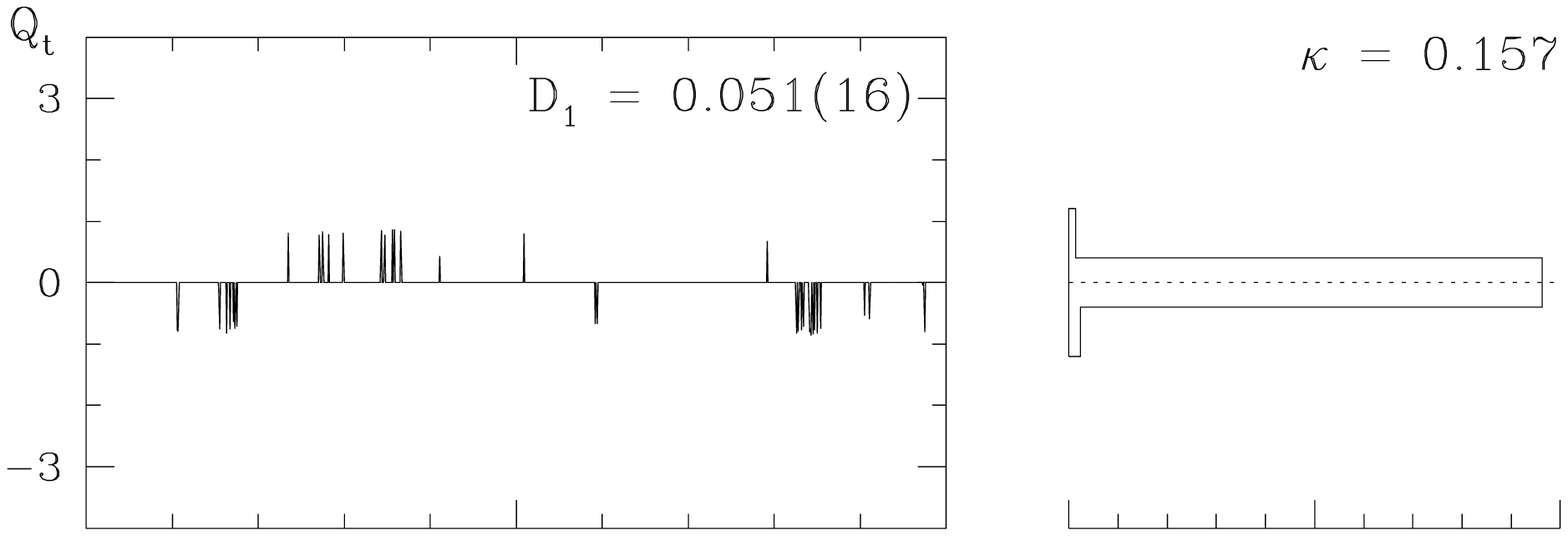}
\qpic{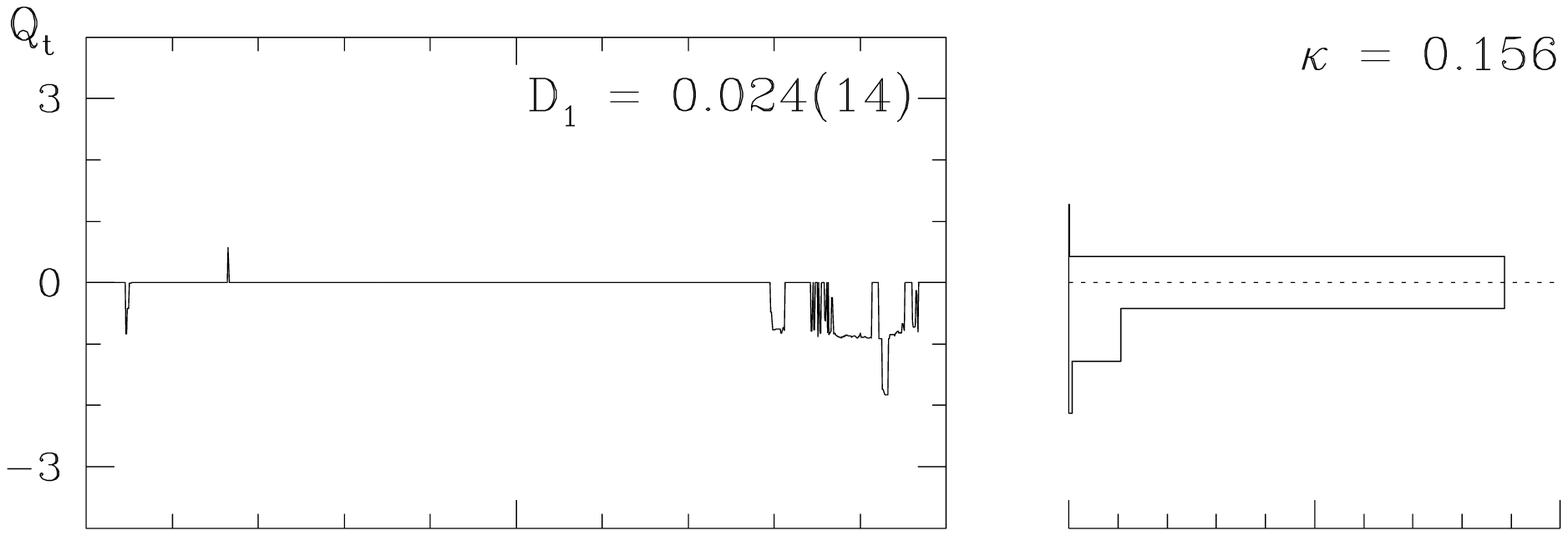}\hspace{1cm}\qpic{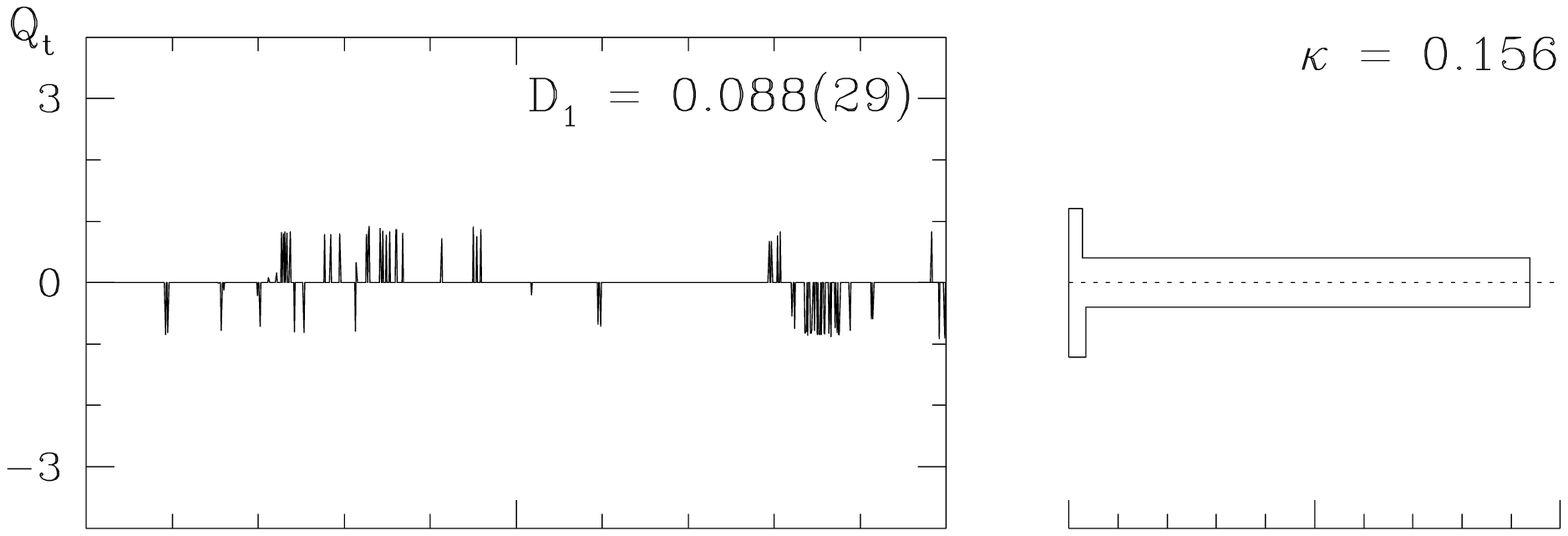}
\qpic{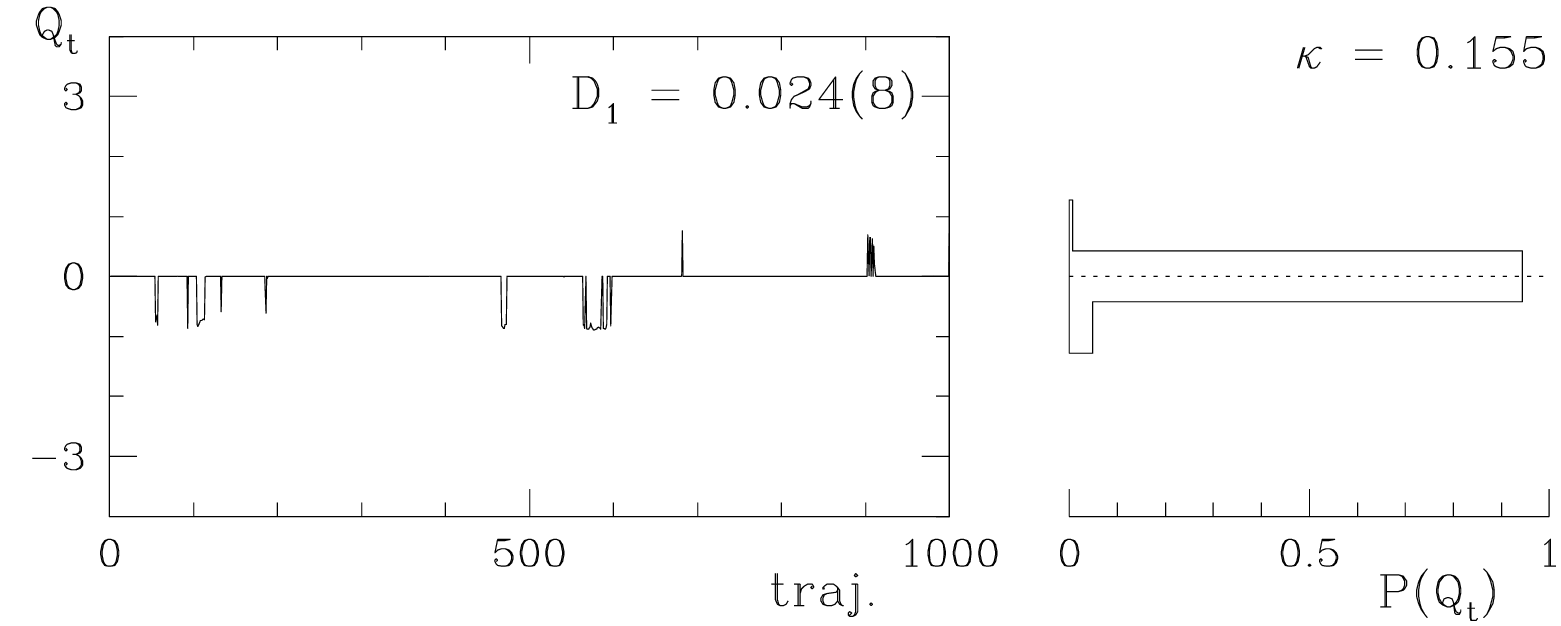}\hspace{1cm}\qpic{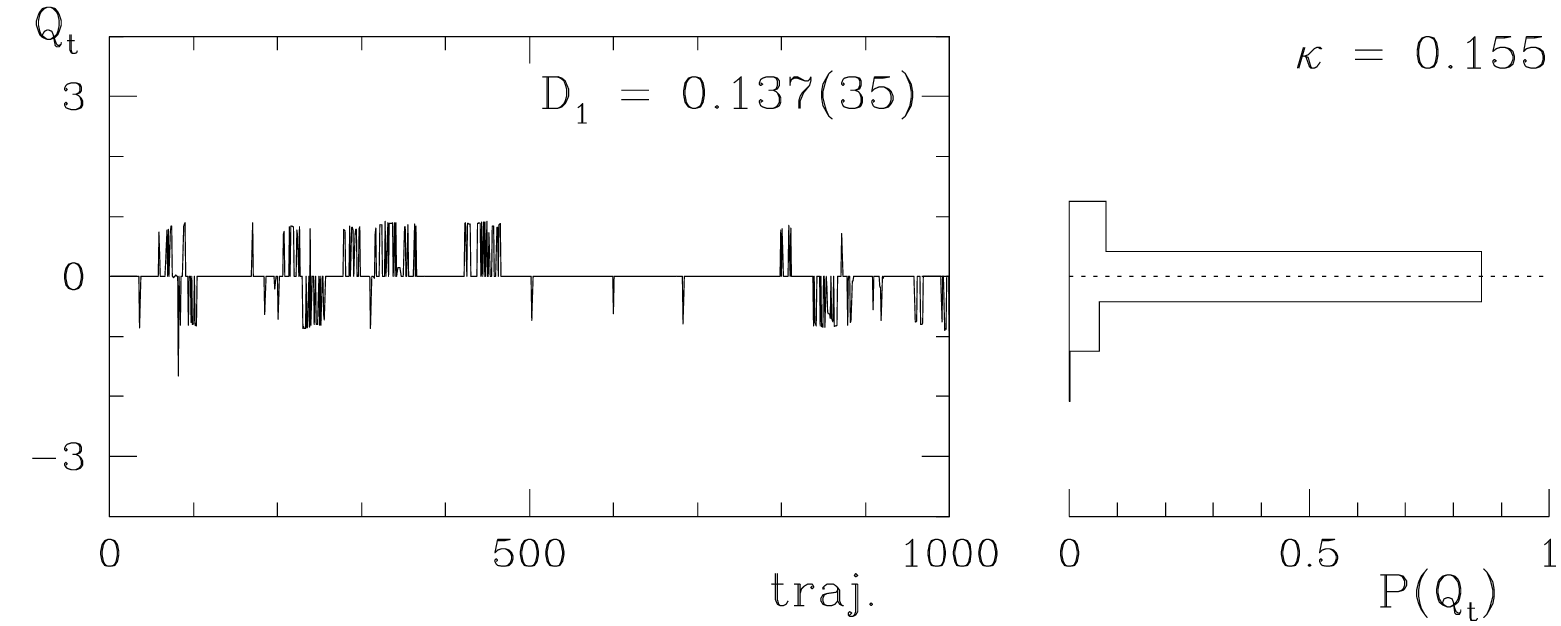}

\vspace*{-5ex}

\caption{Comparison of standard with tempered HMC. The tempering data shown 
are from the run with 6 ensembles in Table 1.}
\end{figure*}

\section{PARALLEL TEMPERING}

In standard Monte Carlo simulations one deals with one parameter set 
$\lambda$ and generates a sequence of configurations $C$. The set $\lambda$ 
here includes $\beta$, $\kappa$, the leapfrog time step and the number of 
time steps. $C$ comprises the gauge field and the pseudo fermion field.

In the parallel tempering approach \cite{HN,EM} one simulates $N$ 
ensembles ($\lambda_i; C_i),\,i = 1, \dots, N$ in a
single combined run.  Two steps alternate: (1) update of $N$
configurations in the standard way, (2) exchange of configurations by
swapping pairs. Swapping of a pair of configurations means
\[
((\lambda_i; C_i), (\lambda_j; C_j)) \rightarrow \left\{
\begin{array}{ll}
\!((\lambda_i; C_j), (\lambda_j; C_i)), & \!\mbox{if acc.} \\
\!((\lambda_i; C_i), (\lambda_j; C_j)), & \!\mbox{else}
\end{array}
\right.
\]
with the Metropolis acceptance condition 
\begin{equation}
P_{\rm swap}(i,j) = \min\left( 1, e^{-\Delta H} \right)
\label{Pswap}
\end{equation}
\[
\Delta H = 
H_{\lambda_i}(C_i) + H_{\lambda_j}(C_j) -
H_{\lambda_i}(C_j) - H_{\lambda_j}(C_i).
\]
Since after swapping both ensembles remain in equilibrium, the swapping 
sequence can be freely chosen. Swapping neighboring pairs and 
proceeding in parameter direction towards criticality has turned out to be
most advantageous.

\section{SIMULATION DETAILS}

We used the standard Wilson action for the gauge and the fermion fields 
and worked on an $8^4$ lattice.  Our HMC program applied the
standard\pagebreak

\noindent
conjugate gradient inverter with even/odd preconditioning. The trajectory 
length was always~1. The time steps were adjusted to get acceptance
rates of about 70\%. In all cases we generated 1000 
trajectories (plus 50--100 trajectories for thermalization).

$Q_t$ was measured by the field-theoretic method after 50 cooling steps of
Cabibbo-Marinari type.

\begin{table*}[t]
\caption{Mobilities $D_1$ (see (\ref{D1})) on the $8^4$ lattice at
$\beta=5.6$. The swap acceptance rates (\ref{Pswap}) achieved were
about 82\% for $\Delta\kappa = 0.00025$ and about 63\% for
$\Delta\kappa = 0.0005$.}
\begin{tabular*}{\textwidth}{l@{\extracolsep{\fill}}l@{\extracolsep{\fill}}lll}
\hline
$\kappa$   & Standard HMC & \multicolumn{3}{c}{Tempered HMC} \\
\cline{1-1}  \cline{2-2}    \cline{3-5}
  &              & 7 ensembles  & 6 ensembles & 21 ensembles \\
  &              & $0.156 \leq \kappa \leq 0.1575$ 
                 & $0.155 \leq \kappa \leq 0.1575$
                 & $0.15 \leq \kappa \leq 0.16$ \\
  &              & $\Delta\kappa = 0.00025$ 
                 & $\Delta\kappa = 0.0005$ 
                 & $\Delta\kappa = 0.0005$ \\
\hline
0.1500 &     0.285(34)  &              &                &     0.668(41)\\[0.5ex]
0.1550 &     0.024(8)   &              &      0.137(35) &     0.106(26)\\
0.1555 &                &              &      0.141(35) &     0.093(24)\\
0.1560 &     0.024(14)  &     0.053(22)&      0.088(29) &     0.076(19)\\     
0.1565 &                &     0.032(13)&      0.079(23) &     0.057(17)\\
0.1570 &     0.002(2)   &     0.018(6) &      0.051(16) &     0.035(11)\\
0.1575 &                &     0.003(2) &      0.033(11) &     0.026(10)\\
0.1580 &     0.002(2)   &              &                &     0.013(6) \\[0.5ex]
0.1600 &     0          &              &                &     0 \\
\hline
\end{tabular*}
\vspace{-3mm}
\end{table*}

\section{RESULTS}

We have run several tempered HMC simulations in the
quenched approximation (tempering in $\beta$) and with dynamical
fermions (tempering in $\kappa$, at fixed $\beta=5.5$ and $\beta=5.6$).  For 
comparison also standard HMC simulations have been performed.

Figure 1 shows a typical comparison of results. We find that with 
tempering considerably more topologically nontrivial configurations occur 
and the histograms of $Q_t$ become more symmetrical. 

In standard runs $Q_t$ frequently stayed for quite some time near 1 or near
$-1$, while with tempering this never occurred. The standard run at 
$\kappa = 0.156$ shown in Figure 1, where $Q_t$ gets trapped in this way for 
about 200 trajectories, provides an example of this. Such observations have
also been made on large lattices \cite{SESAM,RB2}.  

While a correlation analysis cannot be carried out with the size of our 
samples, some quantitative account of the improvement by tempering is
possible using the mean of the absolute change of $Q_t$,
called mobility in \cite{SESAM},
\begin{equation}
D_1 = \frac{1}{N_{\rm traj}} \sum_{i=1}^{N_{\rm traj}}
\left| Q_t(i) - Q_t(i - 1) \right |\;.
\label{D1}
\end{equation}
We give results for $D_1$ in Table~1. The $\kappa$-values of
our tempered runs include the ones used by SESAM \cite{SESAM}.
The considerable gain by tempering is obvious from Figure~1 and Table~1.

For comparisons in Table~1 one should, in addition to the individual 
errors at each of the $\kappa$-values (which are not small), take the 
overall behavior within the particular runs into account. 
Comparing the tempering results for 7 and 6
ensembles it appears that adding intermediate points (not shown in Table~1)
in this case leads to less improvement than extending the range to lower 
$\kappa$-values does. The comparison of the results for 21 and 6 ensembles 
possibly indicates that extending the range to very large $\kappa$-values 
(with no transitions in standard simulations) 
affects the results at lower $\kappa$. In any case our data show that by 
a clever choice of range and distances of the parameter values the number 
of parameter points can be strongly reduced.

\section{CONCLUSIONS}

Parallel tempering considerably enhances tunneling between different
sectors of topological charge and generates samples with more symmetrical
charge distributions than can be obtained by standard HMC. The histograms 
also get slightly broader or even become nontrivial thanks to this 
technique.

The enhancement of tunneling indicates an improvement of decorrelation 
also for other observables. More satisfactory histograms are important
for topologically sensitive quantities. Both of these features make parallel 
tempering an attractive method for large-scale QCD simulations. The method 
is particularly economical when several parameter values have to be studied 
anyway.  

A potential problem which remains to be studied is that for a given parameter 
set the swap acceptance rate (\ref{Pswap}) decreases for increasing lattice
volume \cite{UKQCD}. In this context possibly the choice of 
range and distances of the parameter values gets more important. We therefore 
plan to extend our investigations to larger lattices.

\section*{ACKNOWLEDGEMENTS}

We would like to thank M.~M\"uller-Preussker for support. We are indebted to 
R.~Burkhalter for informations on CP-PACS results. Our
simulations were done on the CRAY T3E at Konrad-Zuse-Zentrum f\"ur
Informationstechnik Berlin.

\end{document}